# Dynamic Rules for Decoherence

Thomas L. Cooper
813 Alder Street
Gilroy, CA 95020

In orthodox quantum theory, decoherence is presumed to be caused by observation. In this paper, the idea of replacing observation, as the cause of decoherence, with rules derived from the dynamics of the system, is addressed. Such rules determine the timing of decoherence and the states in the mixture afterward. For instance, energy conservation during decoherence, for each possible transition, leads to a timing rule. Exponential decay and ergodic behavior follow directly from the dynamic rules as do Boltzman's postulate of equally probable micro-states and the Pauli rate equations. Ergodic behavior in mesoscopic systems is predicted and those predictions are strikingly similar to behavior observed in at least two laboratories.

## I. INTRODUCTION

For this paper, "decoherence" is defined as any transition between states of a system that cannot be described by a transformation in Hilbert space[1]. As far as I can tell, this definition is consistent with all others. It has been chosen to emphasize the view that a law of decoherence is a necessary, but *missing*, part of quantum theory.

The theory of Ghiraldi, Rimini and Weber (GRW) [1] embodies a similar view concerning the need for new law to describe decoherence [2,3]. In GRW, however, Schrodinger's equation is modified to account for decoherence. In the work presented here, Schrodinger's equation is not modified and the law governing decoherence is added.

In his book, von Neumann [4] argues that two fundamentally different processes occur in quantum mechanics. One is Schrodinger evolution: the continuous transformation of one state into another, uniquely determined, state. The other, caused by observation, is the discontinuous (instantaneous) transition from a specific initial state into one of several different final states. Each final state corresponds to a possible outcome of the observation. Observation, however, is never defined. It is debatable whether or not von Neumann's view is still a majority view among physicists. Nevertheless, we shall refer to it as the orthodox view.

Since the final state in a transformation is uniquely determined by the initial state, an instantaneous transition that begins with a specific initial state and ends with one of several different final states cannot be described by a transformation. Accordingly, in the orthodox view, observation, albeit undefined, causes decoherence, as decoherence is defined above.

Several theoretical approaches to the problem of decoherence, other than the orthodox view, have been proposed during the last few decades. Besides GRW, they include the many universes view [5,6], decoherent or consistent histories [7,8,9], Bohmian mechanics [10,11] and environmental interaction [12,13,14]. Some of these theories use observation as the cause of decoherence and some do not [9].

In this paper, it is *not* assumed that observation causes decoherence. Decoherence is assumed to occur when the system attains appropriate dynamic conditions. Decoherence, itself, is a statistical process that occurs in sub-macroscopic (molecular and sub-molecular) systems. As such, we might expect decoherence to be a *missing*, statistically irreversible, *sub-macroscopic* process that would lead directly to statistical irreversibility in macroscopic, thermodynamic systems. After all, statistical behavior appears in both quantum systems and thermodynamic systems. It is argued, in this paper, that a proper theory of decoherence accounts for both and provides a reductionist link from the sub-macroscopic realm to thermodynamic irreversibility.

## II. TWO-STEP DECOHERENCE

When modifying quantum theory, it must be kept in mind that the orthodox theory has been immensely successful in accounting for natural phenomena. For this reason, neither a modification of Schrodinger's equation nor another mode of continuous evolution is proposed here. *Decoherence, it is assumed, results from discontinuities in Schrodinger evolution.* It follows that we can think of decoherence as a sequence of two-step processes. Each two-step process consists of a Schrodinger transformation followed by an *instantaneous* decohering transition.

Other theories assume a non-zero time for decoherence [13,14], so it is important, in our discussions, to distinguish between instantaneous decohering-transitions and non-instantaneous decohering-transitions. Therefore, an *instantaneous* decohering-transition is called a "stochastic transition". Further, measurements resulting in non-zero decoherence-times [15] are assumed to be measurements of the duration of Schrodinger transformations prior to, and between, stochastic transitions.

The detection of randomly timed transitions and transitions from a specific initial state to one of several different final states suggests the existence of transitions that are not transformations in Hilbert space, which, in turn, suggests the likelihood of natural law, in addition to Schrodinger's equation, to govern quantum transitions. Further, there are many examples of randomly timed transitions from a specific initial state to one of several different final states that could reasonably be construed as taking place in the absence of observation. Consider, for example, the transitions occurring in sodium in the presence of a magnetic field. Selection rules allow transitions from either of the $M_J = \pm 1/2$ states in the $3^2P_{1/2}$ sub-shell to either of the $M_J = \pm 1/2$ states making up the $3^2S_{1/2}$ sub-shell. That is, each completely specified initial state makes randomly-timed transitions to different final states.

---

[1] The solutions to Schrodinger's equation are transformations in Hilbert space. Sometimes we refer to a solution to Schrodinger's equation as a "Schrodinger transformation".

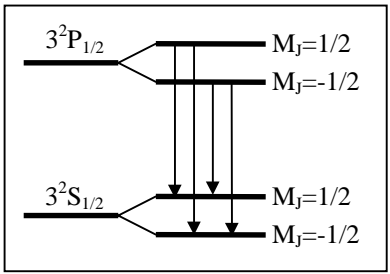

FIG.1. Energy-level diagram for sodium, illustrating randomly-timed transitions from the same initial state to different final states.

It is easy to imagine that transitions, like the ones in sodium occur in the absence of observation. Undoubtedly, that is the way most physicists think of these transitions. To the contrary, it is the orthodox view, requiring observation for an outcome to be realized, that creates conceptual difficulties.

The set of possible states, following a stochastic transition, is called the "final mixture". It is assumed that the final mixture is a subset of at least one basis spanning the Hilbert space for the system under consideration. Further, we assume that the relative probability for each stochastic transition is given, in the usual way, by Born's probability rule.

In writing this paper, it was assumed that the reader would possess a working knowledge of density matrix theory. Results from density matrix theory are often used without proof. To review density matrix theory, the reader may consult one of a large number of sources, including [16].

Let $W_-(t_1)$ and $W_+(t_1)$ be density matrices immediately preceding and immediately following, respectively, a stochastic transition occurring at time $t_1$. Using $W_-(t_1)$ and $W_+(t_1)$, we can summarize our assumptions, so far, as follows.

*Hypothesis 1.0: For a stochastic transition from a pure state that occurs at time, $t_1$, the density matrix immediately prior to the stochastic transition is given by,*

$$W_-(t_1) = |\psi(t_1)\rangle\langle\psi(t_1)|, \qquad (1a)$$

*and the density matrix immediately following the stochastic transition, (final density matrix) is given by*

$$W_+(t_1) = \sum_n |\gamma_n^1\rangle p_n \langle\gamma_n^1|, \qquad (1b)$$

where 
$$|\psi(t_1)\rangle = \sum_n |\gamma_n^1\rangle\langle\gamma_n^1|\psi(t_1)\rangle \qquad (1c)$$

and 
$$p_n^1 = |\langle\gamma_n^1|\psi(t_1)\rangle|^2. \qquad (1d)$$

The superscript, 1, inside Dirac vectors, indicates that a single stochastic transition has occurred since the initial state was specified. The need for keeping track of the number of stochastic transitions will become clear soon.

*It is critical to remember that the system makes a stochastic transition to only one state in the final mixture. The final density matrix is nothing more than a mathematical construct that allows us to keep track of all possible transitions at once.*

Expressions (1), without change, can be used to describe the orthodox view. Collapse, in the orthodox view, is a stochastic transition. This result may be surprising, but it is a good indication that we are on the right track. After all, the orthodox view is consistent with a large number of experimental results.

If expressions (1) are the same for both the orthodox view and the view presented here, what, then, differentiates the two? The rules used to determine the timing of stochastic transitions and the rules used to determine the states in the final mixture differentiate the description of decoherence discussed here from the orthodox view. For the orthodox view, these rules must contain at least an element of "free choice". An observer may choose what and when he will observe. For the orthodox view, these two choices, and the projection postulate[2], are rules that can determine the timing of collapse and final mixture following collapse. The next hypothesis defines an alternative approach for finding rules to determine timing and final states.

*Hypothesis 2.0: The timing for stochastic transitions and states in the final mixture following a stochastic transition are determined uniquely by the dynamics of the system under consideration.*

It is useful to consider, qualitatively, the ramifications of combining the first two hypotheses before proceeding with the detailed arguments. The first hypothesis implies that stochastic transitions are Markov processes. Therefore, a sequence of stochastic transitions should adopt the properties of a Markov chain including relaxation and ergodic [17] behavior. Both properties play important roles in describing irreversible physical processes, but both have been awkward, at best, to derive from first principles. With the orthodox view, producing a Markov chain would require a sequence of several observations. On the other hand, if the stochastic transitions were to occur spontaneously, as required by hypothesis two, Markov chains would emerge naturally. It will be shown that, by requiring energy to be conserved for each stochastic transition, both exponential decay and ergodic behavior can be easily deduced.

---

[2] The projection postulate says that an observation leaves the system in a state corresponding to the observed eigenvalue.

Earlier, we argued that two-step decoherence was consistent with the existence of more than one transition from a specific initial state. In the theorem that follows, we see that the complementary property is also true. That is, if only one transition from a specific initial state is allowed, then that transition is a Schrodinger transformation.

*Theorem 1.0: All transitions between pure states[3] are Schrodinger transformations.*

*Proof*: Consider a system in a state, $|\psi(0)\rangle$, at $t = 0$. Let a stochastic transition occur at $t = t_1$. It is assumed that no other stochastic transitions occur in the interval $0 < t < t_2$, where $t_2 > t_1$. If $W_+(t_1) = W_-(t_1)$, then the transition from $|\psi(0)\rangle$ to $|\psi(t)\rangle$, where $0 < t < t_2$, is a continuous solution to Schrodinger's equation. Then, if it can be shown that $W_+(t_1) = W_-(t_1)$, the theorem is proven.

For $W_-(t_1)$ and $W_+(t_1)$ to represent pure states, each mixture must contain a single state. Then, from (1b),

$$W_+(t_1) = |\gamma_n^1\rangle\langle\gamma_n^1|. \tag{2}$$

From (1c), $\quad |\psi(t_1)\rangle = |\gamma_n^1\rangle. \tag{3}$

Substituting (3) into (2) gives

$$W_+(t_1) = |\psi(t_1)\rangle\langle\psi(t_1)|.$$

Then, using (1a), we see that $W_+(t_1) = W_-(t_1)$, and the theorem is proven. QED.

Care must be taken in applying theorem one. Many atomic and subatomic transitions appear to be transitions between pure states, *but are not*. From (1c) we see that the Schrodinger transformation of the initial state causes a superposition to evolve. That superposition could include the initial state. Since a stochastic transition can be to any state in the superposition, it could be to the initial state. *By recognizing the occurrence of stochastic transitions "back" to the initial state, we can envision nearly any atomic or subatomic transition as being a stochastic transition.*

---
[3] By a "transition between pure states" we mean that all transitions from a specific initial state are to the same final state.

Expressions (1) can be thought of as describing the stochastic transition from a pure state, originating from an initial pure state at $t = 0$, to a mixture at $t = t_1$. If subsequent stochastic transitions occur at times, $t = t_k$, where $k = 1, 2, \ldots$, then for $k > 1$, these transitions will be from one mixture to another mixture. *A stochastic transition from a mixture occurs when at least one of the states in the mixture undergoes a stochastic transition.*

Using the theory of density matrices, it can be shown that the generalization of (1) to a sequence of stochastic transitions, originating from a pure state, $|\psi(0)\rangle$, at $t = 0$, is given by:

$$W_-(t_1) = U(t_1)|\psi(0)\rangle\langle\psi(0)|U^+(t_1), \tag{4a}$$

$$W_+(t_k) = \sum_n |\gamma_n^k\rangle p_n^k(t_k)\langle\gamma_n^k|, \tag{4b}$$

$$W_-(t_{k+1}) = U(\Delta t_k)W_+(t_k)U^+(\Delta t_k), \tag{4c}$$

$$p_n^1(t_1) = \left|\langle\gamma_n^1|U(t_1)|\psi(0)\rangle\right|^2, \tag{4d}$$

and $p_n^{k+1}(t_{k+1}) = \sum_m \left|\langle\gamma_n^{k+1}|U(\Delta t_k)|\gamma_m^k\rangle\right|^2 p_m^k(t_k), \tag{4e}$

where $U(t)$ is the unitary operator satisfying Schrodinger's equation, $U(0)=1$, $\Delta t_k = t_{k+1} - t_k$, $|\gamma_n^k\rangle$ denotes a state in the final mixture following the $k^{th}$ stochastic transition and $p_n^k(t_k)$ is the probability that the system is in the state, $|\gamma_n^k\rangle$, immediately following the $k^{th}$ stochastic transition. The probabilities, $p_n^k(t_k)$, are sometimes called population probabilities. Further, we denote any basis that contains the $k^{th}$ final mixture by $\{|\gamma_n^k\rangle\}$.

### III. CONSERVATION OF ENERGY

A violation of conservation of energy has never been observed in atomic, nuclear or sub-nuclear phenomena.

*Hypothesis 3.0: Energy is conserved exactly across stochastic transitions.*

Hypothesis three holds, whether the system is open or closed. In closed systems, energy is also conserved between stochastic transitions. Then, in a closed system, for a state, $|\gamma_n^k\rangle$, to be included in the final mixture following a stochastic transition at $t = t_k$, it is necessary that

$$\langle \gamma_n^k | H | \gamma_n^k \rangle = \langle \psi(0) | H | \psi(0) \rangle,$$

The converse formulation of conservation of energy is more useful for our purpose.

*Theorem 2.0: Let a closed system, in a state $|\psi(0)\rangle$ at $t = 0$, incur a sequence of stochastic transitions at times $t_k$. If, for any system state, $|\gamma_n^k\rangle$, such that*

$$\langle \gamma_n^k | H | \gamma_n^k \rangle \neq \langle \psi(0) | H | \psi(0) \rangle, \quad (5)$$

*then,* $\quad \langle \gamma_n^1 | U(t_1) | \psi(0) \rangle = 0,$ (6a)

*and, for all k,* $\quad \langle \gamma_n^{k+1} | U(\Delta t_k) | \gamma_m^k \rangle = 0,$ (6b)

*Proof:* By hypothesis three, in a closed system, states contained in $\{|\gamma_n^k\rangle\}$ such that (5) holds cannot be part of the final mixture. So for these states, $p_n^1(t_1) = 0$. Substituting $p_n^1(t_1) = 0$ into (4d), results in equation (6a).

To complete the proof we use the principle of mathematical induction. If $p_m^k(t_k) = 0$ in the RHS of (4e), except for terms for which $\langle \gamma_m^k | H | \gamma_m^k \rangle = \langle \psi(0) | H | \psi(0) \rangle$, then to assure transitions that conserve energy, $p_n^{k+1}(t_{k+1}) = 0$ for states such that $\langle \gamma_n^{k+1} | H | \gamma_n^{k+1} \rangle \neq \langle \gamma_m^k | H | \gamma_m^k \rangle$. This result, in turn, requires that $\langle \gamma_n^{k+1} | U(\Delta t_k) | \gamma_m^k \rangle = 0$.

In the RHS of $p_n^2(t_2) = \sum_m \left| \langle \gamma_n^2 | U(\Delta t_1) | \gamma_m^1 \rangle \right|^2 p_m^1(t_1)$, $p_m^1(t_1) = 0$ for those terms for which $\langle \gamma_m^1 | H | \gamma_m^1 \rangle \neq \langle \psi(0) | H | \psi(0) \rangle$. Then, by the principle of mathematical induction, the theorem is proven. QED.

Expression (5) asserts that $|\gamma_n^k\rangle$ does not correspond to a transition that conserves energy. Equations (6), assure that $|\gamma_n^k\rangle$ is excluded from final mixtures. The time that the first stochastic transition occurs, then, can be found by solving equation (6a) for $t_1$ using, for $|\gamma_n^1\rangle$, all states, contained in $\{|\gamma_n^1\rangle\}$, such that (5) holds. If there are no such solutions, then, no stochastic transition occurs. A similar procedure is used to calculate $t_k$ for $k>1$.

Solutions to (6) may not always correspond to stochastic transitions. If a solution corresponds to a transition to a pure state, then, by theorem one, that transition is not a stochastic transition.

For some systems, $t_1 = 0$ is a solution to (6). For a stochastic transition to occur at $t = 0$, would require that we specify the initial state, $|\psi(0)\rangle$, at a discontinuity. To avoid this difficulty, the solution $t_1 = 0$ is prohibited from corresponding to a stochastic transition. Similarly, for a subsequent stochastic transition, it is necessary that $\Delta t_k > 0$ in (6b).

If there are several solutions to equations (6), then the earliest non-zero time, that does not correspond to a transition between pure states, is the time that the stochastic transition occurs. The question of what happens if (6) is identically zero is not addressed in this paper.

In this paper, equations (6) are not solved. Instead, the implications of hypotheses one through four are investigated by assuming the existence of a solution. The reason for taking this direction is that the existence of a solution to equations (6) restricts the Hamiltonian, thus introducing a significant new topic for investigation. On the other hand, some new and important results can be obtained without knowledge of a specific solution to (6).

### IV. THE ORIGIN OF RANDOM TIMING IN THE DECAY OF UNSTABLE SYSTEMS

Theorem two tells us that the time a stochastic transition occurs can be calculated by solving equations (6) for time. Finding the time that a stochastic transition occurs, this way, raises a critical issue. The decay of an unstable system occurs with random timing, but solutions to equations (6) are not random. They are unambiguously determined by the equation. How, then, can the time of the decay of an unstable system be random?

Imagine an atom initially in an excited state. Assume that the atom makes a stochastic transition at a time $t_1$. The stochastic transition causes the atom to decay or to return to its initial state according to the probabilities of (1d). If the stochastic transition returns the atom to its initial (excited) state, the atom will make a second stochastic transition at time $2t_1$. Again, a final state is randomly selected from among the initial state and possible states corresponding to decay. *Decay occurs after a random number of transitions back to the excited state.*

The importance of the process, discussed above, is that it permits randomly timed decay even if the interval between each stochastic transition is determined unambiguously. To assure that random decay can occur from unambiguously timed stochastic transitions, however, the excited state must be included in the final mixture for every stochastic transition in the sequence. Accordingly, for the excited state to be in each final mixture, the final density matrix for each transition must commute with final density matrix for the subsequent transition.

*Hypothesis 4.0: At least for some systems that incur a sequence of stochastic transitions,*

$$[W_+(t_{k+1}), W_+(t_k)] = 0 \text{ for all } k.$$

We call sequences that satisfy hypothesis four, "commuting sequences" and sequences that do not, "non-commuting".

Hypothesis four contains a hedge. It is not suggested that all sequences of stochastic transitions are commuting sequences. The possibility is left open that there are, as yet unknown, conditions that determine whether on not a sequence is commuting. For now, in cases where hypothesis four is needed, it is assumed that whatever conditions may exist are satisfied.

Hypothesis four is a necessary but *insufficient* condition for including the initial state in a sequence of stochastic transitions. In addition to hypothesis four, the initial state must be included in every basis that contains the first final mixture. Note that choosing an initial state from a basis containing the first final mixture is always possible for commuting sequences, but is not necessarily possible, otherwise.

*Theorem 2.0 For a commuting sequence of stochastic transitions the time between any two successive transitions is the same. That is*

$$t_{k+1} = t_k + \Delta t_o. \qquad (7)$$

*Proof*: For commuting sequences, the superscripts, *k*, can be dropped from expressions (4). Then the bases containing the states in every final mixture in the sequence can be denoted by $\{|\gamma_n\rangle\}$ instead of $\{|\gamma_n^k\rangle\}$. To obtain (7), we note that, for commuting sequences, the solution to (6b), $\langle\gamma_j|U(\Delta t_k)|\gamma_n\rangle = 0$, where $\langle\gamma_j|H|\gamma_j\rangle \neq \langle\gamma_n|H|\gamma_n\rangle$, is independent of *k*. That is, $\Delta t_k = \Delta t_o$. QED.

## V. EXPONENTIAL DECAY IN OPEN SYSTEMS

The decay of an unstable system, such as an excited atom, in an open system should result in an exponential form with time and with a well-known time constant. Equations (6) hold for closed systems only. For open systems, however, hypothesis three still holds. Then in an open system energy enters and leaves the system *between* stochastic transitions. Other than these ideas, however, we do not have a theory of stochastic transitions for open systems. Nonetheless, it is possible to deduce exponential decay, in open systems, by making an additional, physically reasonable, assumption.

Let $|\gamma_i\rangle$ denote the initial state of a system and let that system be an excited atom. Let each of the possible final states, $|\gamma_f\rangle$, describe the atom in its ground state with a photon resulting from the decay. Between stochastic transitions lthe photon leaves the system.

To show that exponential decay results from stochastic transitions, we begin by examining (4e). We assume that the stochastic transitions form commuting sequences. Then (4e) has the form,

$$p_n(t_{k+1}) = \sum_m |\langle\gamma_n|U(\Delta t_k)|\gamma_m\rangle|^2 p_n(t_k). \qquad (8)$$

Each term in (4e) can be interpreted as either an emission term or an absorption term. Now then, we make the physically reasonable, but otherwise unjustified, assumption that we can describe the decay of an excited atom in an open system by dropping all of the absorption terms from (8). That is, for all states, $|\gamma_f\rangle$,

$$p_f(t_k) = 0 \qquad (9)$$

Substituting expressions (9) into (8) and setting $n = i$,

$$p_i(t_{k+1}) = |\langle\gamma_i|U(\Delta t_o)|\gamma_i\rangle|^2 p_i(t_k). \qquad (10)$$

Further, since all final mixtures are contained in a basis, $\{|\gamma_n\rangle\}$, then, by completeness,

$$1 = \sum_m |\langle\gamma_m|U(\Delta t_0)|\gamma_i\rangle|^2. \qquad (11)$$

We are concerned only with the population probabilities at times $t_1$ and $t_k = t_1 + (k-1)\Delta t_o$. Then, by theorem two, all the terms in (11) vanish except those for which $\langle\gamma_m|H|\gamma_m\rangle = \langle\gamma_i|H|\gamma_i\rangle$. Rearranging (11) results in

$$|\langle\gamma_i|U(\Delta t_0)|\gamma_i\rangle|^2 = 1 - \sum_{m \neq i} |\langle\gamma_m|U(\Delta t_0)|\gamma_i\rangle|^2. \qquad (12)$$

Note that the states for which $m \neq i$ are the final states that, earlier, had been denoted by $|\gamma_f\rangle$. Substituting (12) into (10) gives,

$$p_i(t_{k+1}) - p_i(t_k) = -p_i(t_k) \sum_{m \neq i} |\langle\gamma_m|U(\Delta t_0)|\gamma_i\rangle|^2. \qquad (13)$$

We define
$$\frac{1}{\tau} = \sum_{m \neq i} \frac{|\langle \gamma_m | U(\Delta t_0) | \gamma_i \rangle|^2}{\Delta t_0}. \quad (14)$$

Then, dividing both sides of (13) by $\Delta t_0$ and substituting (14) into the result gives

$$\frac{p_i(t_{k+1}) - p_i(t_k)}{\Delta t_0} = -\frac{p_i(t_k)}{\tau}. \quad (15)$$

Assuming that $\frac{\Delta t_0}{\tau} \ll 1$, expression (15) can be approximated by

$$\frac{dp_i(t_k)}{dt_k} = -\frac{p_i(t_k)}{\tau}. \quad (16)$$

Equation (16) has the well-known solution,

$$p_i(t_k) = e^{-t_k/\tau}. \quad (17)$$

The idea of using multiple "collapses" to explain exponential decay has been proposed before. See references [18], [19] and references therein, especially reference 13 in [19].

## VI. ERGODIC BEHAVIOR

From (4e), for commuting sequences, in closed systems,

$$p_n(t_{k+1}) = \sum_m |\langle \gamma_n | U(\Delta t_k) | \gamma_m \rangle|^2 p_m(t_k) \quad (18)$$

Expression (18) defines a Markov chain with $|\langle \gamma_n | U(\Delta t_0) | \gamma_m \rangle|^2$ playing the part of the part of the transition probability for both quantum theory and the theory of Markov chains. Using the same arguments that were used to derive (13), it is straightforward to show that

$$p_n(t_{k+1}) - p_n(t_k) = \sum_{m \neq n} |\langle \gamma_n | U(\Delta t_o) | \gamma_m \rangle|^2 p_m(t_k)$$
$$- \sum_{m \neq n} |\langle \gamma_m | U(\Delta t_o) | \gamma_n \rangle|^2 p_n(t_k) \quad (19)$$

It is clear from the form of (19) that stationary solutions, independent of the initial system state, are possible. That is, solutions, independent of initial conditions, exist such that

$$p_n(t_{k+1}) = p_n(t_k) \equiv p_n.$$

We define the transition rate, $w_{mn}$, as,

$$w_{mn} = \frac{|\langle \gamma_m | U(\Delta t_0) | \gamma_n \rangle|^2}{\Delta t_0}. \quad (20)$$

Then, substituting (20) into (19) gives

$$\frac{p_n(t_{k+1}) - p_n(t_k)}{\Delta t_o} = \sum_{m \neq n} [w_{nm} p_m(t_k) - w_{mn} p_n(t_k)]. \quad (21)$$

The stationary solutions for (21) are well-known and are given by

$$w_{mn} p_n = w_{nm} p_m. \quad (22)$$

for all $m$, $n$ such that $\langle \gamma_n | H | \gamma_n \rangle = \langle \gamma_m | H | \gamma_m \rangle$.

The expressions (22) are a form of the principle of detailed balance. If, in addition to (22), $|\langle \gamma_m | U(\Delta t_o) | \gamma_n \rangle| = |\langle \gamma_n | U(\Delta t_o) | \gamma_m \rangle|$, then it follows, from (20), that

$$w_{mn} = w_{nm}. \quad (23)$$

Combining (23) with (22) yields Boltzman's postulate of equally probable micro-states, used in the derivation of the Maxwell-Boltzman distribution [20]. That is,

$$p_n = p_m \quad (24)$$

for all $m$, $n$ such that $\langle \gamma_n | H | \gamma_n \rangle = \langle \gamma_m | H | \gamma_m \rangle$.

The results in this section are intriguing. I believe it is the first time that Boltzman's postulate of equally probable micro-states has been derived from principles that are applicable to situations other than thermodynamic equilibrium. Further, we see that statistical behavior is not limited to systems with a large number of degrees of freedom. Experimental evidence supporting these results can be found in references [21] and [22]. These results are discussed in detail in section XII of this paper.

Next, consider expression (21) for the case that

$$\frac{p_n(t_{k+1}) - p_n(t_k)}{\Delta t_0} \ll w_{nm}.$$

In this case, we can find a good approximation for $p_n(t_k)$ by solving the set of differential equations,

$$\frac{dp_n}{dt} = \sum_{m \neq n} [w_{nm} p_m(t) - w_{mn} p_n(t)].$$

These are the well-known Pauli rate equations [23].

## VII. CLOSED MESOSCOPIC SYSTEMS

In the previous two sections, purely stochastic descriptions of exponential decay and ergodic behavior were developed. In neither section was the dynamic behavior (Schrodinger evolution) between stochastic transitions considered. For many systems, a purely stochastic description is all that is needed. In others, however, both the stochastic transitions and Schrodinger evolution between stochastic transitions are important. The next theorem provides us with the ability to describe a system both stochastically and dynamically, over time, during a sequence of commuting stochastic transitions.

*Theorem 4.0: For a commuting sequence of stochastic transitions, in a closed system, the population probability, $p_n(t)$, for each final-mixture state, $|\gamma_n\rangle$, is given by*

$$p_n(t) = |\langle \gamma_n |U(t)|\psi(0)\rangle|^2 \text{ for } 0 \leq t \leq t_1 \quad (25a)$$

and $$p_n(t) = \sum_m |\langle \gamma_n |U(t-t_k)|\gamma_m\rangle|^2 p_m(t_k) \quad (25b)$$

for $t_k \leq t \leq t_{k+1}$

*Proof*: Expression (25a) is the usual expression for $p_n(t)$. Expression (25b) follows from density matrix theory. That is, for $t_k \leq t \leq t_{k+1}$,

$$p_n(t) = \langle \gamma_n |U(t-t_k)W_+(t_k)U^+(t-t_k)|\gamma_m\rangle \quad (26)$$

Substituting (4b) into (26) yields (25b). QED.

It is, of course, also possible to derive expressions similar to (25) for a system during a sequence of non-commuting transitions. The additional generality, however, comes at the cost of more complexity, so we do not consider the general case here.

Let $|\gamma_2\rangle$ denote an excited atom and $|\gamma_1\rangle$ denote the atom in its ground state with a photon. Further, let $\langle\gamma_1|H|\gamma_1\rangle = \langle\gamma_2|H|\gamma_2\rangle$. Assume that the other atom/photon states, $|\gamma_j\rangle$, are such that $\langle\gamma_j|H|\gamma_j\rangle \neq \langle\gamma_1|H|\gamma_1\rangle$. Now assume that the transition rates among all states, except between $|\gamma_1\rangle$ and $|\gamma_2\rangle$, are small enough that they can be ignored when solving Schrodinger's equation, but large enough to cause a stochastic transition when equations (6) are satisfied. That is when $\langle\gamma_j|U(\Delta t_o)|\gamma_1\rangle = \langle\gamma_j|U(\Delta t_o)|\gamma_2\rangle = 0$. With the above assumptions, the system behaves like a two level system between stochastic transitions.

The exact solution to Schrodinger's equation for a two-level system is well known and can be found in many textbooks, including the one by Sakarai [24]. With some obvious notational changes,

$$|\langle\gamma_1|U(t-t_k)|\gamma_2\rangle|^2 = \sin^2 \Omega(t-t_k), \quad (27a)$$

from which it follows that

$$|\langle\gamma_1|U(t-t_k)|\gamma_1\rangle|^2 = \cos^2 \Omega(t-t_k), \quad (27b)$$

for $t_k < t < t_{k+1}$, where $k \geq 0$, $\Omega$ is twice the Rabi frequency, and, by convention, $t_o = 0$. Noting that, because equations (6) must be satisfied at $t = t_k$,

$$p_1(t_k) + p_2(t_k) = 1. \quad (28)$$

Substituting expressions (27) and (28) into (25) gives

$$p_1(t) = p_1(t_k)\cos^2 \Omega(t-t_k) \\ + [1-p_1(t_k)]\sin^2 \Omega(t-t_k) \quad (29)$$

for $t_k < t < t_{k+1}$, $k \geq 0$ and $t_o = 0$.

The key parameter to use for interpreting expression (29) is the ratio $\frac{\Omega\Delta t_o}{2\pi}$. If $\frac{\Omega\Delta t_o}{2\pi} = \frac{n}{2}$, where $n$ is an integer, then by theorem one, there are no stochastic transitions and undamped Rabi oscillations occur. If $\frac{\Omega\Delta t_o}{2\pi} = \frac{1}{4}$ expression (29) becomes stationary immediately after the first stochastic transition with a value, $p_1(t) = 1/2$, for $t > \frac{\pi}{2\Omega}$.

For every value of $\frac{\Omega\Delta t_o}{2\pi} \neq \frac{n}{2}$, $p_1(t)$ either reaches a stationary value of $1/2$ in a finite time, or approaches it asymptotically. That is, expression (29) predicts that ergodic behavior can occur in a two level system and further predicts the conditions for ergodic behavior and the stationary values of the population probabilities. Note that the stationary values, predicted by (29), agree with the equal probabilities predicted by expression (24).

Results from a calculation, using (29), for $\frac{\Omega\Delta t_o}{2\pi} = 0.43$, and $\frac{\Omega\Delta t_o}{2\pi} = 0.38$, are shown in figure two. Note in figure 2(a), that the Rabi oscillations are damped and that the population probabilities appear to be approaching an asymptotic value of $1/2$. In figure 2(b), the stationary value of $1/2$ is reached after just two stochastic transitions.

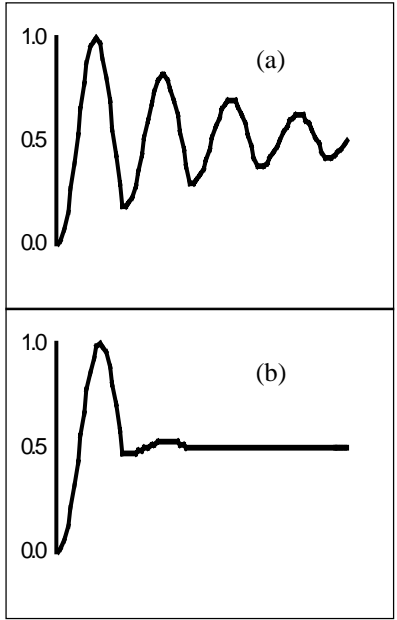

FIG 2. Suppressed Rabi oscillations in a two level system, calculated from expression (29). The ordinate is $p_1(t)$ for an atom initially in the excited state. The curve in (a) is for $\frac{\Omega \Delta t_o}{2\pi} = 0.43$ and (b) is for $\frac{\Omega \Delta t_o}{2\pi} = 0.38$.

Evidence of ergodic behavior in mesoscopic systems can be found in the results of groups in Paris [21] and at NIST [22]. Both results show suppression of Rabi oscillations. Further, stationary population probabilities of $1/2$ occur for each level. The reader is encouraged to compare figure two to figures in references [21] and [22].

A discussion of the importance of the results in references [21] and [22], as well as an explanation of the results that is different from the explanation proposed here, can be found in a paper by Bonifacio, Olivares, Tombesi and Vitali [25]. Other investigators [26,27,28] have also addressed the NIST results by assuming that the suppression of Rabi oscillation can be properly described by exponential damping.

## VIII. SUMMARY AND CONCLUSION

Decoherence has been described using a few simple assumptions. A summary of these assumptions, grouped into the four hypotheses, follows.

Hypothesis 1.0
1. Schrodinger evolution and isolated discontinuities in Schrodinger evolution, called stochastic transitions, can describe all quantum transitions.
2. The states in the final mixture, resulting from a stochastic transition, are contained in at least one basis spanning the Hilbert space for the system.
3. The probability for a stochastic transition into a particular final state is given by Born's probability rule.

Hypothesis 2.0:
1. The states in the final mixture, resulting from a stochastic transition, are determined by system dynamics.
2. The timing of stochastic transitions is determined by system dynamics.

Hypothesis 3.0:
Energy is conserved for each stochastic transition.

Hypothesis 4.0:
There are systems that incur sequences of stochastic transitions for which $[W_+(t_{k+1}), W_+(t_k)] = 0$.

From the foregoing assumptions, we have been able to derive some surprising but seemingly requisite results. For instance, the origin of irreversible behavior in closed systems has long been a central and controversial issue for physics [29,30,31]. The problem centers on the perceived lack of any stochastic law governing the behavior of the sub-macroscopic world. It has been argued here, that a law of decoherence is the missing sub-macroscopic stochastic law. As evidence for the correctness of this idea, it has been shown here, that the assumptions, listed above, regarding such a law, lead directly and unequivocally to ergodic behavior in sub-macroscopic systems. Ergodic behavior is statistically irreversible behavior. That is, the expectation value of any time-independent observable will attain, in time, a stationary value.

An underlying assumption for statistical mechanics has been that it addresses only systems with a large number of degrees of freedom. The results, presented here, do not require that assumption and predict ergodic behavior in closed systems with a small number of degrees of freedom. The predicted behavior is remarkably similar to behavior observed in at least two mesoscopic systems.

Another common assumption regarding statistical mechanics is also apparently contradicted by the results presented here. That assumption is that there is a such a thing as purely "classical statistical mechanics". Classical mechanics provides no mechanism for ergodic behavior.

Boltzman's postulate of equal probabilities for micro-states is derived in a straightforward manner. *Not only has Boltzman's postulate for equally probable micro-states, been derived, but also direct experimental evidence for the postulate has been found in mesoscopic systems.*

Throughout this work, seemingly disparate ideas regularly showed remarkable consistency. Nonetheless, the view presented in this paper has serious shortcomings.
1. A rule for determining the states in the final mixtures was not proposed.
2. Equations (6) were not solved or even shown to have solutions[4].
3. It is not clear that the violation of Bell's inequality can be maintained in the presence of stochastic transitions[5].

---

[4] Equations (6) can be solved using perturbation theory.
[5] In subsequent work, I have found that there is at least one uniquely determined final mixture for which Bell's inequality is violated in the presence of stochastic transitions.

Even with the shortcomings, I submit that the results presented here warrant further investigation.